      [1999/12/01 v1.4c Il Nuovo Cimento]
\documentclass{cimento}

\usepackage{graphicx} 
\title{Measurement of heavy-flavour production in ALICE}
\author{A.~Dainese\from{ins:x}, for the ALICE Collaboration}
\instlist{\inst{ins:x} INFN - Laboratori Nazionali di Legnaro, Legnaro (Padova), Italy}

\newcommand{\sqrtsNN}{\sqrt{s_{\scriptscriptstyle{{\rm NN}}}}}

\newcommand{\gev}{\mathrm{GeV}}
\newcommand{\tev}{\mathrm{TeV}}

\newcommand{\cm}{\mathrm{cm}}

\newcommand{\mum}{\mathrm{\mu m}}

\newcommand{\PbPb}{\mbox{Pb--Pb}}

\newcommand{\pp}{\mbox{proton--proton}}

\newcommand{\pt}{p_{\rm t}}

\newcommand{\dEdx}{{\rm d}E/{\rm d}x}
\newcommand{\dNdy}{{\rm d}N_{\rm ch}/{\rm d}y}

\newcommand{\ccbar}{\mbox{$\mathrm {c\overline{c}}$}}
\newcommand{\bbbar}{\mbox{$\mathrm {b\overline{b}}$}}

\newcommand{\Dz}{\mbox{$\mathrm {D^0}$}}

\newcommand{\Jpsi} {\mbox{J\kern-0.05em /\kern-0.05em$\psi$}\xspace}

\PACSes{\PACSit{25.75.-q}\PACSit{14.65.Dw}}

\begin{document}

\maketitle

\begin{abstract}
The ALICE experiment, currently in the commissioning phase, 
will study nucleus--nucleus and proton--proton collisions at the 
CERN Large Hadron Collider (LHC). We review the ALICE heavy-flavour 
physics program and present a selection of results on the expected
performance for the case of proton--proton 
collisions.
\end{abstract}

\section{Introduction}
\label{intro}

The main physics goal of the 
ALICE experiment~\cite{alicePPR1} is the study of nucleus--nucleus
collisions at the LHC, with a centre-of-mass energy per
nucleon--nucleon collision $\sqrtsNN=5.5~\tev$ for the Pb--Pb system, 
in order to investigate the properties of QCD matter at energy densities of 
up to several hundred times the density of atomic nuclei. Under 
these conditions
a deconfined state of quarks and gluons is expected to be formed.

The measurement of open charm and open beauty production allows to 
investigate the mechanisms of heavy-quark production, propagation and
 ha\-dro\-ni\-za\-tion
in the hot and dense medium formed in high-energy nucleus--nucleus collisions.
Heavy-quark production measurements in proton--proton collisions 
at the LHC energy of $\sqrt{s}=14~\tev$, 
besides providing the necessary baseline for
the study of medium effects in nucleus--nucleus collisions, are interesting 
{\it per se}, as a test of QCD in a new energy domain.

\section{Heavy-flavour production at LHC energies}

Heavy-quark pairs ($\rm Q\overline Q$) 
are expected to be produced in 
primary partonic scatterings with large virtuality $Q^2>(2m_{\rm Q})^2$.
Therefore, 
the baseline production cross sections in nucleon--nucleon collisions can be 
calculated in the framework of perturbative QCD
(pQCD). 
For the estimate of baseline production yields in nuclear collisions,
scaling of the yields with the average number
of inelastic nucleon--nucleon collisions is usually assumed.
The expected 
$\ccbar$ and $\bbbar$ production yields for pp collisions at 
$\sqrt{s}=14~\tev$ are 0.16 and 0.0072, respectively~\cite{notehvq}. 
For the 5\% most central Pb--Pb collisions at $\sqrtsNN=5.5~\tev$ 
the expected yields are 115 and 4.6, respectively.
These numbers, assumed as the baseline for ALICE simulation studies, 
are obtained from pQCD calculations at NLO~\cite{hvqmnr}, including 
the nuclear modification of the parton distribution functions 
(PDFs)~\cite{EKS98}
in the Pb nucleus 
(details on the choice of pQCD parameter values and PDF sets can be found 
in~\cite{notehvq}).
Note that the predicted yields have large uncertainties, of about a factor 2,
estimated by varying the values of the calculation parameters.
An illustration of 
the theoretical uncertainty bands for the D and B 
meson cross sections
will be shown in sections~\ref{opencharm} and \ref{openbeautyele}, 
along with the expected 
sensitivity of the ALICE experiment.

\section{Heavy-flavour detection in ALICE}
\label{exp}

The ALICE experimental setup, described in detail in~\cite{alicePPR1,zampolli},
allows the detection
of ${\rm D}$ and ${\rm B}$ mesons in the high-multiplicity environment 
of central \PbPb~collisions at LHC energy, where a few thousand 
charged particles might be produced per unit of rapidity. 
The heavy-flavour capability of the ALICE detector is provided by:
\begin{itemize}
\item Tracking system; the Inner Tracking System (ITS), 
the Time Projection Chamber (TPC) and the Transition Radiation Detector (TRD),
embedded in a magnetic field of $0.5$~T, allow track reconstruction in 
the pseudorapidity range $-0.9<\eta<0.9$ 
with a momentum resolution better than
2\% for $\pt<20~\gev/c$ 
and a transverse impact parameter\footnote{The transverse impact parameter,
$d_0$, is defined as the distance of closest approach of the track to the 
interaction vertex, in the plane transverse to the beam direction.} 
resolution better than 
$60~\mum$ for $\pt>1~\gev/c$ 
(the two innermost layers of the ITS, $r\approx 4$ and $7~\cm$, 
are equipped with silicon pixel 
detectors).
\item Particle identification system; charged hadrons are separated via 
$\dEdx$ in the TPC and via time-of-flight measurement in the 
Time Of Flight (TOF) detector; electrons are separated from charged 
hadrons in the dedicated
Transition Radiation Detector (TRD), and in the TPC; 
muons are identified in the muon 
spectrometer covering the pseudo-rapidity range $-4<\eta<-2.5$~\cite{stocco}. 
\end{itemize}

Simulation studies~\cite{alicePPR2}
have shown that ALICE has good potential to carry out
a rich heavy-flavour physics programme. The main analyses in preparation 
are:
\begin{itemize}
\item Open charm (section~\ref{opencharm}): fully reconstructed hadronic decays 
$\rm D^0 \to K^-\pi^+$, $\rm D^+ \to K^-\pi^+\pi^+$,
$\rm D_s^+ \to K^-K^+\pi^+$ (under study), $\rm \Lambda_c^+ \to p K^-\pi^+$ (under study) in $|\eta|<0.9$.
\item Open beauty (sections~\ref{openbeautyele} and~\ref{openbeautymu}): 
inclusive single leptons ${\rm B\to e}+X$ 
in $|\eta|<0.9$ and ${\rm B\to\mu}+X$ in $-4<\eta<-2.5$; inclusive displaced
charmonia ${\rm B\to J/\psi\,(\to e^+e^-)}+X$ (under study).
\item Quarkonia (covered in~\cite{stocco}): $\rm c\overline c$ (J/$\psi$, 
$\psi^\prime$) and $\rm b\overline b$ ($\Upsilon$, 
$\Upsilon^\prime$, $\Upsilon^{\prime\prime}$) states 
in the ${\rm e^+e^-}$ ($|\eta|<0.9$) and $\mu^+\mu^-$ ($-4<\eta<-2.5$) 
channels.
\end{itemize} 
For all simulation studies, a multiplicity $\dNdy=4000$--$6000$
was assumed for central \mbox{Pb--Pb} collisions.
In the following, we report the results corresponding to the 
expected statistics collected by ALICE per LHC year: 
$10^7$ central (0--5\% $\sigma^{\rm inel}$) \mbox{Pb--Pb} events at
$\mathcal{L}_{\rm Pb-Pb}=10^{27}~\cm^{-2}{\rm s}^{-1}$
and $10^9$ pp events at 
$\mathcal{L}_{\rm pp}^{\rm ALICE}=5\times 10^{30}~\cm^{-2}{\rm s}^{-1}$,
in the barrel detectors; the forward muon arm will collect
about 40 times larger samples of muon-trigger events
(i.e.\, $4\times 10^8$ central \mbox{Pb--Pb} events).

\section{Charm reconstruction}
\label{opencharm}

Among the most promising channels for open charm detection are the 
$\rm D^0 \to K^-\pi^+$ ($c\tau\approx 120~\mum$, branching ratio 
$\approx 3.8\%$) and $\rm D^+ \to K^-\pi^+\pi^+$ ($c\tau\approx 300~\mum$, 
branching ratio $\approx 9.2\%$) decays. The detection strategy
to cope with the large combinatorial background from the underlying event 
is based on the selection of displaced-vertex topologies, i.e. separation 
from the primary vertex of
the tracks from the secondary vertex 
and good alignment between the reconstructed D meson momentum 
and flight-line~\cite{alicePPR2,elena}. 
An invariant-mass analysis is used to extract the raw signal 
yield, to be then corrected for detector acceptance and 
for selection and reconstruction efficiency.
The accessible $\pt$ range for the $\Dz$ 
is $1$--$20~\gev/c$ in \mbox{Pb--Pb} and 
$0.5$--$20~\gev/c$ in pp, 
with statistical errors better than 15--20\% at high $\pt$~\cite{alicePPR2}. 
Similar capability 
is expected for the $\rm D^+$~\cite{elena}, 
though at present the
statistical errors are estimated only in the range $1<\pt<8~\gev/c$.
The systematic errors 
(acceptance and efficiency corrections, 
centrality selection for Pb--Pb) are expected to be smaller than 20\%.

\begin{figure}[!t]
  \begin{center}
  \includegraphics[width=0.49\textwidth]{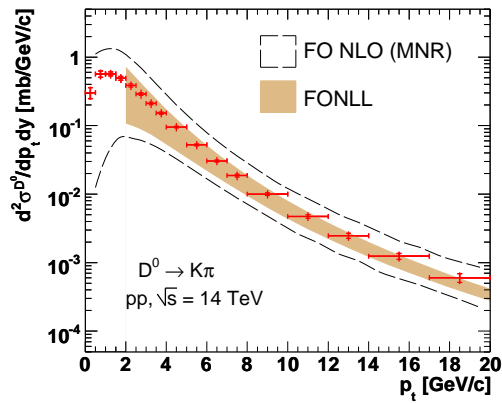}
  \caption{Sensitivity on d$^2\sigma^{\rm D^0}/$d$\pt$d$y$,
           in pp at 14~TeV, compared to 
             NLO pQCD predictions from the MNR~\cite{hvqmnr} and 
             FONLL~\cite{fonll} calculations.
             The inner error bars represent the statistical errors,
             the outer error bars represent the quadratic sum of 
             statistical and $\pt$-dependent systematic errors.
              A normalization error of 5\% is not shown.}
\label{fig:D0ptcmp}
\end{center}
\end{figure}

For the case of pp collisions, the experimental errors on the 
$\pt$-differential cross section are
expected to be significantly smaller than the current theoretical uncertainty 
from perturbative QCD calculations. 
In Fig.~\ref{fig:D0ptcmp} we superimpose the simulated ALICE measurement 
points for the $\Dz$ in pp collisions
to the prediction bands from the MNR fixed-order massive 
calculation~\cite{hvqmnr} and from the FONLL fixed-order next-to-leading log
calculation~\cite{fonll,cacciari}.
The perturbative uncertainty bands were estimated by varying the values of the 
charm quark mass and of the factorization and renormalization scales.
The comparison shows that ALICE will be able  to perform 
a sensitive test of the pQCD predictions for charm production at LHC energy.

\section{Beauty via single electrons}
\label{openbeautyele}

The production of open beauty can be studied by detecting the 
semi-electronic decays of beauty hadrons, mostly B mesons. 
Such decays have a branching ratio of $\simeq 10\%$ 
(plus 10\% from cascade decays ${\rm b\to c} \to e$, that only populate 
the low-$\pt$ region in the electron spectrum).
The main sources of background electrons are: decays of D mesons; 
$\pi^0$ Dalitz decays 
and decays of light vector mesons (e.g.\,$\rho$ and $\omega$);
conversions of photons in the beam pipe or in the inner detector 
layer and pions misidentified as electrons. 
Given that electrons from beauty have average 
impact parameter $d_0\simeq 500~\mum$
and a hard $\pt$ spectrum, it is possible to 
obtain a high-purity sample with a strategy that relies on:
electron identification with a combined $\dEdx$ (TPC) and transition
radiation (TRD) selection;
impact parameter cut to 
reduce the charm-decay component and 
reject misidentified $\pi^\pm$ and $\rm e^{\pm}$
from Dalitz decays and $\gamma$ conversions.
As an example, with $200<d_0<600~\mum$ and $\pt>2~\gev/c$, 
the expected statistics
of electrons from b decays is $8\times 10^4$ for $10^7$ central 
\mbox{Pb--Pb} events, allowing the measurement of electron-level 
$\pt$-dif\-fe\-ren\-tial 
cross section in the range $2<\pt<20~\gev/c$ with statistical errors smaller
than 15\% at high $\pt$. Similar performance figures are expected for 
pp collisions.

\begin{figure}[!t]
  \begin{center}
  \includegraphics[width=0.49\textwidth]{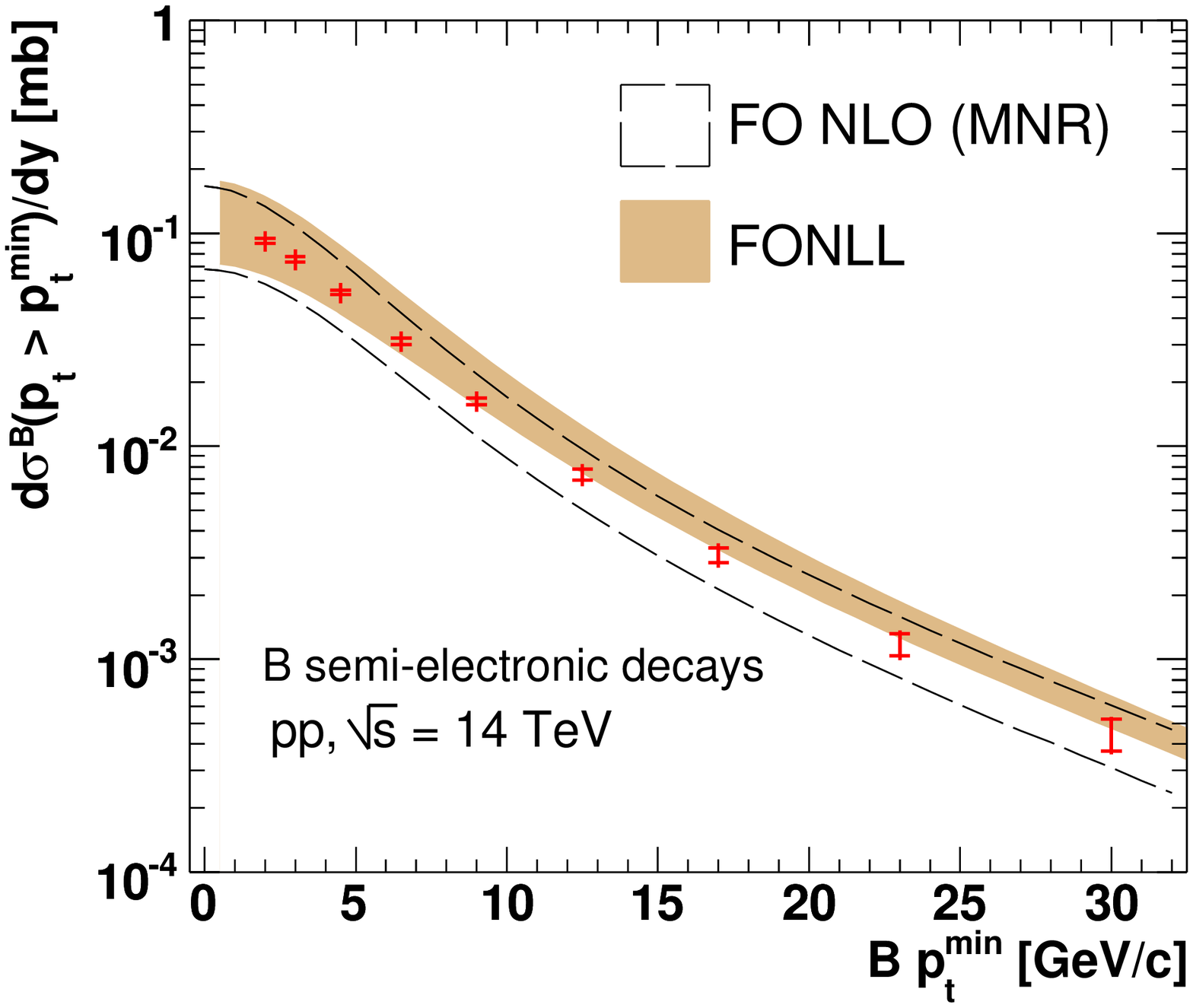}
  \includegraphics[width=.49\textwidth]{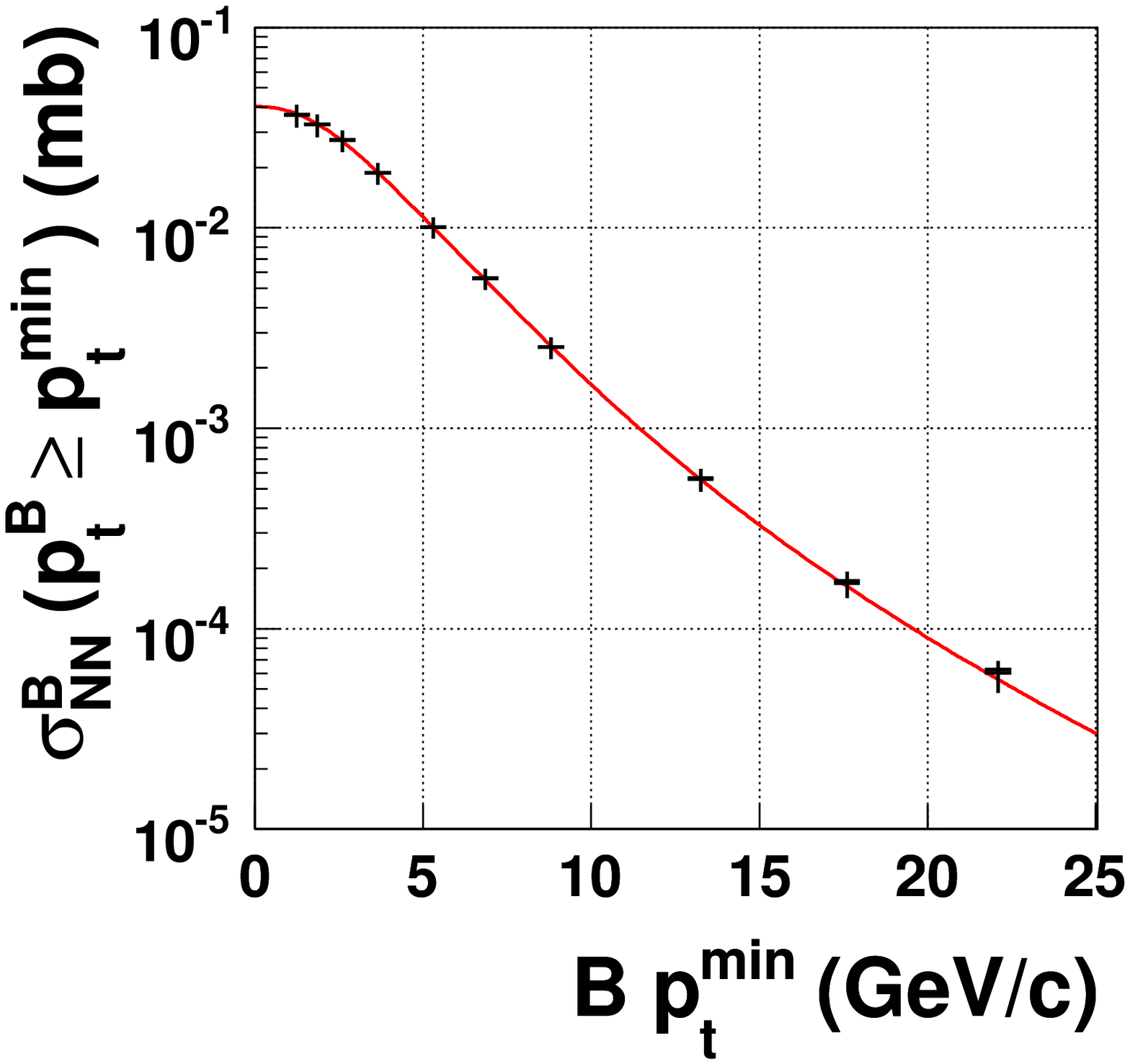}
  \caption{Left: sensitivity on d$\sigma^{\rm B}(\pt>\pt^{\rm min})/$d$y$,
           in pp at 14~TeV, compared to 
             NLO pQCD predictions from the MNR~\cite{hvqmnr} and 
             FONLL~\cite{fonll} calculations;
             error bars are defined as in Fig.~\ref{fig:D0ptcmp}.
          Right: minimum-$\pt$-differential production 
          cross section per nucleon--nucleon collision for B mesons
          with $-4<y<-2.5$ in central Pb--Pb 
           collisions at 5.5~TeV, as expected to be measured 
           from the single-muon data set. 
           Statistical errors (represented by the thickness of the horizontal 
             bars) corresponding to 
            $4\times 10^8$ events
            and $\pt$-dependent systematic errors (vertical bars) 
            are shown. A normalization error of 10\% is not shown.
           The line indicates the input cross section.}
\label{fig:sigmaB}
\end{center}
\end{figure}

Figure~\ref{fig:sigmaB} (left) presents the expected ALICE performance 
for the measurement of the $\pt^{min}$-differential
cross section of B mesons, 
${\rm d}\sigma^{B}(\pt>\pt^{min})/{\rm d}y$ vs. $\pt^{min}$
averaged in the range $|y| < 1$, which can be derived from the electron-level
cross section.
For illustration of the sensitivity in the comparison to pQCD calculations, we 
report in the same figure the predictions and the theoretical uncertainty bands from 
the perturbative calculations in the MNR~\cite{hvqmnr} and FONLL~\cite{fonll,cacciari} approaches.   
It can be seen that the expected ALICE performance for $10^9$ events will provide a meaningful comparison with pQCD 
predictions.

\section{Beauty via single muons}
\label{openbeautymu}

Beauty production can be measured also in the ALICE muon 
spectrometer, $-4<\eta<-2.5$, analyzing the single-muon $\pt$ distribution
and the opposite-sign di-muons invariant mass 
distribution~\cite{alicePPR2}.

The main backgrounds to the `beauty muon' signal are $\pi^\pm$, 
$\rm K^\pm$ and charm decays. The cut $\pt>1.5~\gev/c$ is applied to all
reconstructed muons in order to increase the signal-to-background ratio.
Then, a fit technique allows to extract a $\pt$ distribution of muons 
from B decays.
A Monte Carlo procedure
allows to extract 
B-level cross sections for the data sets (low-mass $\mu^+\mu^-$, 
high-mass $\mu^+\mu^-$, 
and $\pt$-binned single-muon distribution), 
each set covering a different B-meson $\pt>\pt^{\rm min}$ region. 
The simulation results for central Pb--Pb collisions at 
$\sqrtsNN=5.5~\tev$ using only the single muons are 
shown in Fig.~\ref{fig:sigmaB} (right). 
Since only minimal cuts are applied, the reported statistical errors 
(represented by the thickness of the horizontal bars) are very 
small and the high-$\pt$ reach is excellent. Similar performance,
in terms of $\pt$ coverage, is 
expected for pp collisions at $\sqrt{s}=14~\tev$.
The main sources of systematic errors (vertical bars) are: corrections for 
acceptance and efficiency, subtraction of the background 
muons from charged pion and kaon decays, and fit procedure to 
separate the beauty and charm components.

\section{Conclusions}

We presented the expected performance of the ALICE detector 
for the measurement 
of charm and beauty production in $\pp$ collisions at $\sqrt{s}=14~\tev$. 
These measurements will 
provide sensitive tests for perturbative QCD in
a new energy domain. They will also be essential for a comparison with the corresponding measurements
in $\PbPb$ collisions, for example for the investigation of 
c and b quark in-medium energy loss~\cite{hotquarks06}.


\begin{thebibliography}{0}

\bibitem{alicePPR1}
  ALICE Collaboration, Physics Performance Report Vol.~I,  
  CERN/LHCC 2003-049 and \IN{J.~Phys.~G}{30}{2003}{1517}.

\bibitem{notehvq}
  \BY{Carrer N. \atque Dainese A.} 
   ALICE Internal Note, ALICE-INT-2003-019 (2003), 
   arXiv:hep-ph/0311225. 

\bibitem{hvqmnr} 
   \BY{Mangano M.L., Nason P. \atque Ridolfi G.} 
   \IN{Nucl. Phys. B}{373}{1992}{295}.

\bibitem{EKS98} 
  \BY{Eskola K.J., Kolhinen V.J. \atque Salgado C.A.} 
  \IN{Eur.~Phys.~J.~C}{9}{1991}{61}.

\bibitem{zampolli}
  \BY{Zampolli C.} This conference.

\bibitem{stocco}
  \BY{Stocco D.} This conference.

\bibitem{alicePPR2}
  ALICE Collaboration, Physics Performance Report Vol.~II,
  CERN/LHCC 2005-030
  and \IN{J.~Phys.~G}{32}{2006}{1295}.

\bibitem{elena}
  \BY{Bruna E.} \IN{Int. J. Mod. Phys. E}{16}{2007}{2097}

\bibitem{fonll}
  \BY{Cacciari M., Frixione S., Mangano M.L., Nason P. \atque Ridolfi G.}
  \IN{JHEP}{0407}{2004}{033}.

\bibitem{cacciari}
  \BY{Cacciari M.} private communication.

\bibitem{hotquarks06}
  \BY{Dainese A.}  
  \IN{Eur. Phys. J. C}{49}{2007}{135}.

\end{thebibliography}
\end{document}